\documentclass[conference]{IEEEtran}
\IEEEoverridecommandlockouts

\usepackage[colorlinks=false, linkcolor=black, urlcolor=black, citecolor=black, bookmarks=false]{hyperref}
\usepackage[hyphenbreaks]{breakurl}
\usepackage{url}
\usepackage{cite}
\usepackage{amsmath,amssymb,amsfonts}
\usepackage{algorithm}
\usepackage[noend]{algorithmic}

\usepackage{graphicx}
\usepackage{subfigure}
\usepackage{booktabs}
\usepackage{textcomp}
\usepackage{xcolor}
\usepackage{array}
\usepackage{tikz}
\usepackage{etoolbox}
\usepackage{url}
\usepackage{pifont}

\newtheorem{proposition}{Proposition}[section]

\def\BibTeX{{\rm B\kern-.05em{\sc i\kern-.025em b}\kern-.08em
    T\kern-.1667em\lower.7ex\hbox{E}\kern-.125emX}}

\begin{document}

\title{A Microservice Graph Generator with Production Characteristics}

\author{\IEEEauthorblockN{Fanrong Du\IEEEauthorrefmark{2}, Jiuchen Shi\thanks{\IEEEauthorrefmark{2} Fanrong Du and Jiuchen Shi contribute equally to this work.}\IEEEauthorrefmark{2}, Quan Chen\thanks{\IEEEauthorrefmark{3} Quan Chen is the corresponding author.}\IEEEauthorrefmark{3}, Li Li, Minyi Guo}
\IEEEauthorblockA{Shanghai Jiao Tong University}
}

\maketitle

\begin{abstract}

A production microservice application may provide multiple services, queries of a service may have different call graphs, and a microservice may be shared across call graphs.
It is challenging to improve the resource efficiency of such complex applications without proper benchmarks, while production traces are too large to be used in experiments.
To this end, we propose a Service Dependency Graph Generator (DGG) that comprises a \textit{Data Handler} and a \textit{Graph Generator}, for generating the service dependency graphs of benchmarks that incorporate production-level characteristics from traces.
The data handler first constructs fine-grained call graphs with dynamic interface and repeated calling features from the trace and merges them into dependency graphs, and then clusters them into different categories based on the topological and invocation types.
Taking the organized data and the selected category, the graph generator simulates the process of real microservices invoking downstream microservices using a random graph model, generates multiple call graphs, and merges the call graphs to form the small-scale service dependency graph with production-level characteristics.
Case studies show that DGG's generated graphs are similar to real traces in terms of topologies. Moreover, the resource scaling based on DGG's fine-grained call graph constructing increases the resource efficiency by up to 44.8\% while ensuring the required QoS.

\end{abstract}


\section{Introduction}
Microservice architecture has become an emerging paradigm in modern software development, which decomposes the monolithic service into loosely coupled, independently deployable microservices~\cite{7796008,blinowski2022monolithic}.
A service composed of microservices can be denoted as a Directed Acyclic Graph (DAG) where the vertices and edges represent the microservices and call dependencies, respectively~\cite{luo2021,9774016}.
To guarantee the Quality of Service (QoS), the computing resources of each microservice need to be scaled with the load change~\cite{zhang2021sinan,fu2021qos}.

\begin{figure}
  \centering
  \includegraphics[width=0.8\linewidth]{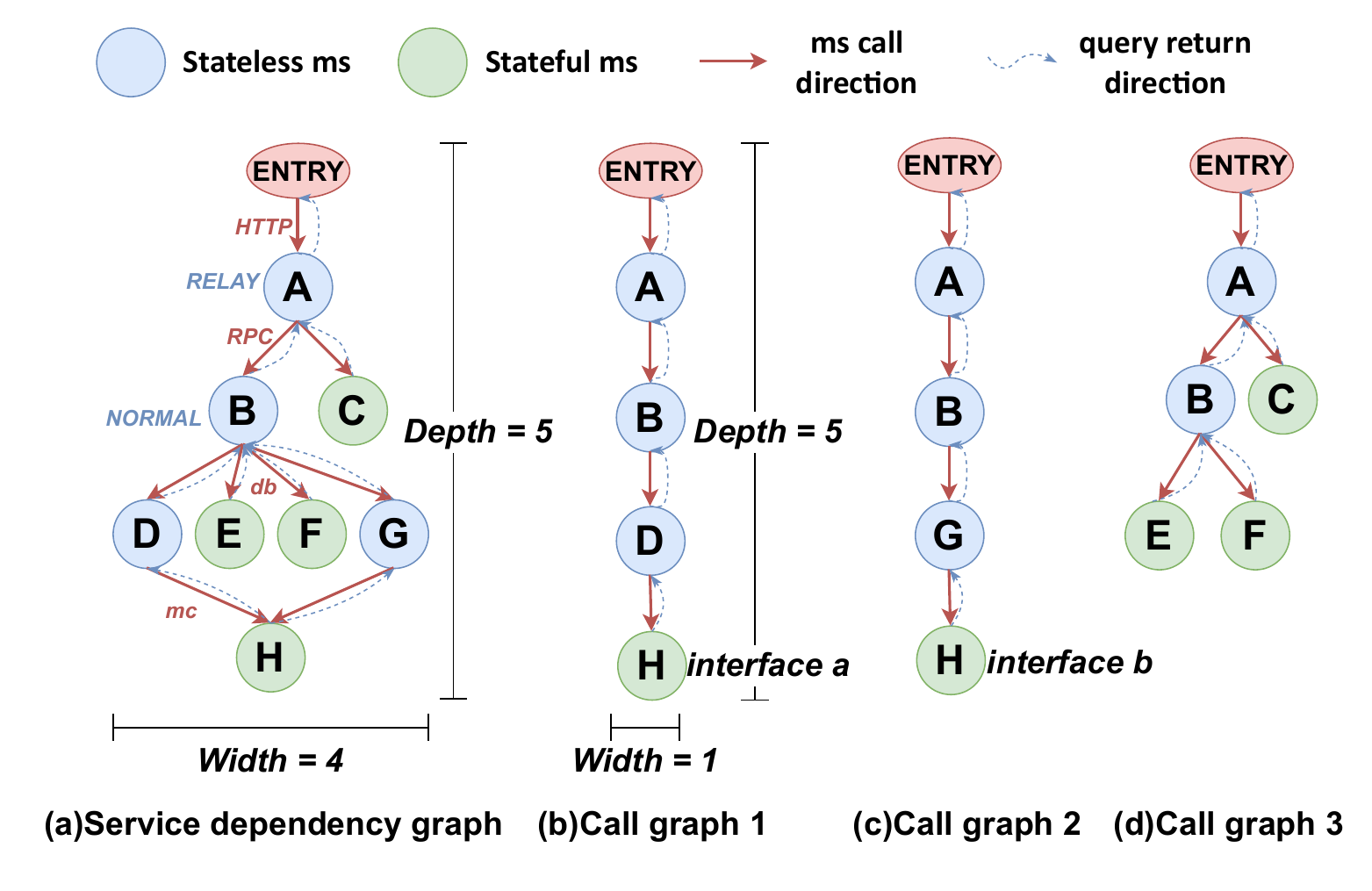}
  \caption{An example service dependency graph and its three call graphs. 
  }
  \label{dependency_graph_example}
 \vspace{-3mm}
\end{figure}

Cloud vendors have open-sourced production microservice traces~\cite{huye2023lifting,luo2021,9774016}. 
The traces show that the service exhibits intricate {\it call graphs} with diverse vertices and edge types. For instance, Fig.~\ref{dependency_graph_example} shows the service dependency graph of a service from the Alibaba trace and its three call graphs.
All the call graphs of a service form its \textit{service dependency graph}.
We can observe that the queries accessing the same service may go through part of the microservices based on user-specified requirements to form different call graphs~\cite{luo2021,9774016}. 
Moreover, there are stateless and stateful microservices~\cite{vayghan2019microservice}, e.g., business logic and databases, and diverse communication modes~\cite{luo2021,9774016}, e.g., Remote Procedure Call (RPC) and HTTP. 
Both stateful and stateless microservices have a set of Application Programming Interfaces (APIs) for upstream microservices to call.
Some microservices are shared by call graphs~\cite{luo2022erms,huye2023lifting}. 
For instance,  microservice H has two interfaces $a$ and $b$ that are called by different call graphs in Fig.~\ref{dependency_graph_example}.

While the traces are too large to be used in research,
it is crucial to develop benchmarks\footnote[1]{A benchmark is a service composed of microservices.} for production microservice applications~\cite{acme_air,musuit,deathstar,Online_Boutique,poster,nodens}.
However, although the benchmarks originating from production services are abstracted on small-scale applications, they lack commonly existing production characteristics like dynamic interfaces~\cite{Online_Boutique,nodens}.
For example, the HotelReservation\cite{deathstar} has 8 call graphs and includes 22 microservices, with a depth of 3 and only 2 microservices featuring dynamic interfaces. However, observed from the real Alibaba trace, it has more than 40 call graphs per service, with the maximum call depth exceeding 15. Moreover, 48.8\% of microservices have more than two interfaces, and a service can include over 500 microservices.
The assistant tools that allow researchers to customize benchmarks~\cite{muBench,sedghpour2023hydragen} also fail to capture realistic production characteristics. 



There are also several automatic tools proposed to generate call graphs with production characteristics~\cite{luo2021,9774016}. However, they are not sufficient for two major defects.

As for the first defect, the trace analysis for microservice characteristics is incomplete. Although existing studies have examined the call graph characteristics of microservices, they still lack a thorough analysis of service dependency graph variation, microservice inter-relationships in call graphs, and various interfaces for individual microservices. These characteristics are crucial for investigating microservice resource scaling, as they can lead to heterogeneous computing resource demand of microservices. 

As for the second defect, the services generated by existing tools have large deviations from reality.
The service dependency graph is generated based on the overall trace statistics, without perceiving the variations between different services.
Moreover, the generated call graphs have little inter-relationship, which cannot reflect microservice sharing features among different call graphs. 
Also, inside a specific call graph, the repeated calling and sibling relationship among microservices are not reflected.
From the perspective of individual microservices, the characteristics of various interfaces are ignored in these tools. 

To address the above issues, we propose a {\it Service Dependency Graph Generator (DGG)} that includes a \textit{Data Handler} and a \textit{Graph Generator} to generate the service dependency graphs of benchmarks that include the production characteristics\footnote[2]{DGG is open-sourced via \url{https://github.com/dufanrong/DGG}.}. 
It is proposed based on our three aspects of new observations from the thorough analysis of production traces.
1) Upstream microservices may have repeated calls to the downstream microservices, and microservices have dynamic interfaces called by different call graphs.
2) Service dependency graphs vary significantly in topology and innovation patterns.
3) Microservices typically invoke a set of child microservices, and the probability of a microservice calling a specific children set is influenced by its sibling microservices.

Based on the above observations, the challenges include capturing repeated invocations and dynamic interfaces, addressing the variability among different service types, as well as representing the invoked microservices as the children set, and considering the effect of sibling microservices.
Therefore, DGG's data handler first constructs fine-grained call graphs from production traces and merges them to form dependency graphs, and then clusters the dependency graphs into different categories based on their topological and invocation features. 
Based on the organized data, DGG's graph generator creates random graph models to represent microservice calls as children set invocations that are influenced by the sibling microservices. 
It then generates simulated call graphs based on these models and merges them to form the final service dependency graph.

We also conduct a case study on using DGG to generate benchmark dependency graphs, and investigate the microservice scaling efficiency under dynamic loads.
In more detail, we use DGG to generate different types of service dependency graphs with associated call graphs, and evaluate the similarities between the generated and real call graphs in terms of the topology and innovation types.
The results show that the call graphs generated by DGG are similar to real-world ones.
Moreover, we investigate the resource allocation efficiency of integrating DGG's fine-grained call graph constructing into microservice resource scaling under dynamic loads.
We use DGG's generated benchmarks, randomly selected benchmarks in the trace, and a realistic benchmark to conduct the investigations. Results show that this strategy can increase the resource efficiency by up to 44.8\% while ensuring the QoS.

The major contributions of this paper are as follows.
\begin{itemize}
    \item \textbf{In-depth analysis of production microservice traces}. The analysis reveals novel observations of production microservices that motivate the design of DGG. 
    \item \textbf{The design of the dependency graph generator that incorporates production characteristics}. DGG first constructs fine-grained graphs from the production trace and clusters them into different categories. Then, DGG generates service dependency graphs for each service type based on novel random graph models.
    \item \textbf{Case studies on benchmark generations and efficient microservice resource scaling}. We validate the similarity between DGG's generated graphs to the real-world ones and the benefits of DGG's fine-grained call graph constructing for efficient microservice resource scaling. 
\end{itemize}

\section{Related Work\label{sec:related_work}}
In this section, we discuss the related work on cloud trace analysis, microservice benchmarks, and call graph generators.

\textbf{Cloud trace analysis:} Some studies analyzed the runtime performance~\cite{amvrosiadis2018diversity,10.1145/3265723.3265742,reiss2012heterogeneity,cortez2017resource} or resource usage~\cite{shi2022characterizing} of lots of types of cloud workloads or production clusters. These works did not aim at the microservice architecture. Moreover, some other studies focused on the characteristics of microservice call graphs~\cite{tian2019characterizing,luo2021,9774016,huye2023lifting,wen2022characterizing}. However, they lacked in-depth exploration of the variations in service dependency graphs, the relationships between microservices in call graphs, and the interfaces of individual microservices.


\textbf{Microservice benchmarks:}
Numerous benchmarks were developed for research on microservice resource scaling~\cite{acme_air,musuit,deathstar,Online_Boutique,poster}, but they failed to reflect the realistic characteristics of real-world microservices~\cite{seshagiri2022sok}. 
Specifically, DeathStarBench~\cite{deathstar} had few call graphs. 
$\mu$suit consisted of four different services, each with only two microservices, whereas production services usually consist of dozens to hundreds of microservices~\cite{10.1145/3190508.3190546,10.1145/3472883.3487003}. 
Moreover, these benchmarks lack rich communication modes of realistic microservices that significantly affect latency and resource management~\cite{google_rest_vs_rpc,dreamfactory_grpc_vs_rest}
For instance, $\mu$suit only adopted gRPC for the inter-microservice communication. 
At last, these benchmarks include little sharing characteristic among microservices, which is commonly existed in production environment~\cite{seshagiri2022sok,luo2022erms}.


Some previous works also proposed assistant tools to support flexible customization of microservice scale, topology, and behaviors by developers~\cite{muBench,sedghpour2023hydragen}. However, they cannot capture production microservice characteristics automatically.

\textbf{Call graph generation tools:}
Luo et al. proposed a call graph generator based on the distribution of production microservices~\cite{luo2021,9774016}. 
However, this generator modeled the overall distribution of all services, and thus cannot capture the variations in different service dependency graphs. 
The generated call graphs are also not inter-connected, making it difficult to construct a complete service dependency graph.
Moreover, this tool failed to consider several production microservice characteristics, including microservice sharing, communication modes, and sibling effects among microservices.
\section{Background and Terminology\label{sec:bg}\label{cg_bg}}
{\bf Service Dependency Graphs and Call Graphs.}
The user queries may go through part of the microservices in the service dependency graph based on user characteristics, forming different call graphs. 
Microservices that are accessed by multiple call graphs are referred to as shared microservices in this paper.
For instance, an e-commerce recommendation service can recommend products based on two filter conditions, including the price and rate microservices. The user queries that select price filtering, rate filtering, and both price and rate filtering will form three different call graphs, respectively. The three call graphs share the price and rate microservices. 

Inside a service dependency graph or call graph, there is an entry microservice for receiving user queries, e.g., nginx~\cite{nginx}. For a specific call, the triggering and called microservices are the upstream and downstream microservices (UM and DM)~\cite{luo2021,9774016}.
An UM typically calls a set of DMs, referred to as the {\it children set}, and the DMs with the same UM are considered as { \it sibling microservices}.
In Fig.~\ref{dependency_graph_example}(d), A is the entry microservice that receives queries via HTTP. 
B and C are sibling microservices, as well as they are DMs of A (which is the UM) and also construct the children set of A.

The topological characteristics of the service dependency graph or call graph mainly encompass depth and width. Depth is defined as the longest path from the entry microservice to any other microservice, while width refers to the maximum microservice number at any given layer of depth.
In Fig.~\ref{dependency_graph_example}(b), the depth of this call graph is 5, and the width is 1.  

{\bf Microservices Types.} There are stateful and stateless microservices in production~\cite{vayghan2019microservice}. 
Stateful ones are typically databases and caching middlewares like mongodb~\cite{mongodb} and memcached~\cite{memcached}.
Stateless ones are mostly related to business logic~\cite{10.1145/3472883.3487003}.
Based on their interactions with downstream microservices, stateless microservices are further categorized into RELAY, LEAF, and NORMAL~\cite{10.1145/3472883.3487003}.
RELAY must have downstream microservices, LEAF no longer calls others, and NORMAL will call others with a certain probability. 


{\bf Communication Modes.}
Microservices mainly communicate 
through Inter-Process Communication (IPC)~\cite{van2002distributed}, Remote Procedure Call (RPC)~\cite{srinivasan1995rpc}, or Message Queue (MQ)~\cite{gan2019open}. 
IPC typically occurs between stateless and stateful microservices~\cite{luo2021}, like calling mongodb and memcached. 
RPC is a type of synchronous communication, in which the requester needs to wait for the reply of the responser with a blocking mode.
MQ is an asynchronous method where microservices communicate via message queues like Kafka~\cite{kreps2011kafka} and RabbitMQ~\cite{ionescu2015analysis}. In this model, there is no need for immediate response, enabling non-blocking communications.

\section{Microservice Trace Analysis}
\label{sec:trace_analysis}
In this section, we analyze the characteristics of the service dependency graphs, call graphs, and call characteristics of microservices in the production clusters of Alibaba and Meta. 
These are the only two open-source microservice traces, representing typical cloud applications (Taobao and Facebook).

\subsection{Overview of Alibaba and Meta Traces\label{trace_overview}}
The Alibaba trace dataset v2022~\cite{alitrace} has over 20 million call graphs involving more than 17,000 microservices across ten clusters over 13 days. 
In this trace, each service is identified by a service ID, while each user query is tracked with a unique trace ID. Each pair of microservice calls is identified by a rpcID, including details about the upstream microservice (UM), the downstream microservice (DM), the type of communication (rpctype), the interface of the downstream microservice invoked, etc.~\cite{luo2021,9774016}.
We randomly select the data of 80\% services in this trace for trace analysis in this section and for evaluating the similarities between DGG's generated and real call graphs in Section~\ref{sec:eval-similarity}.
Moreover, we use the remaining 20\% of the data to validate our trace observations in resource management in Section~\ref{section:case_study_2}, to avoid data leakage.

The Meta trace dataset~\cite{metatrace} features a microservice topology with 18,500 active services and over 12 million service instances. This trace includes information on the service type, call depth, maximum width, the set of downstream microservices (DM set) invoked by each microservice, the number of upstream microservice (UM) invocations to the children set in a single query, and other relevant metrics~\cite{huye2023lifting}.

\subsection{Characterizing Dependency Graphs}
Since the Meta trace~\cite{metatrace} lacks complete dependency graph information, we analyze the characteristics of the Alibaba trace~\cite{alitrace} in this subsection.
We have two major observations.
1) Production service dependency graphs are dynamic over time. These graphs encompass many different call graphs, with their proportions shifting dynamically. 
2) There is significant variation in topological characteristics and microservice invocation patterns across different service dependency graphs. 

\begin{figure}
    \centering
    \subfigure[Total query number over time.]{
        \includegraphics[width=0.44\linewidth]{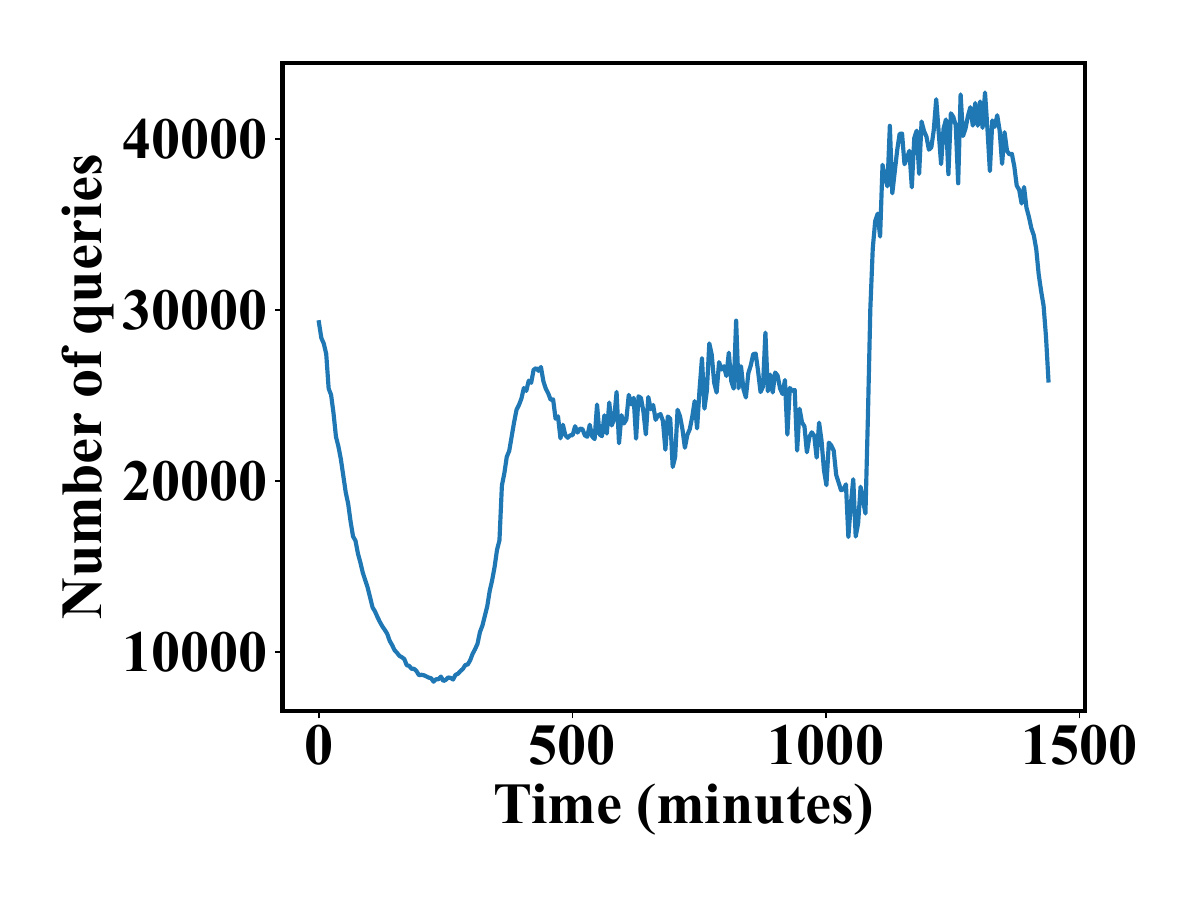}
        \label{total_query}
    }
    \subfigure[Call graph queries over time.]{
        \includegraphics[width=0.44\linewidth]{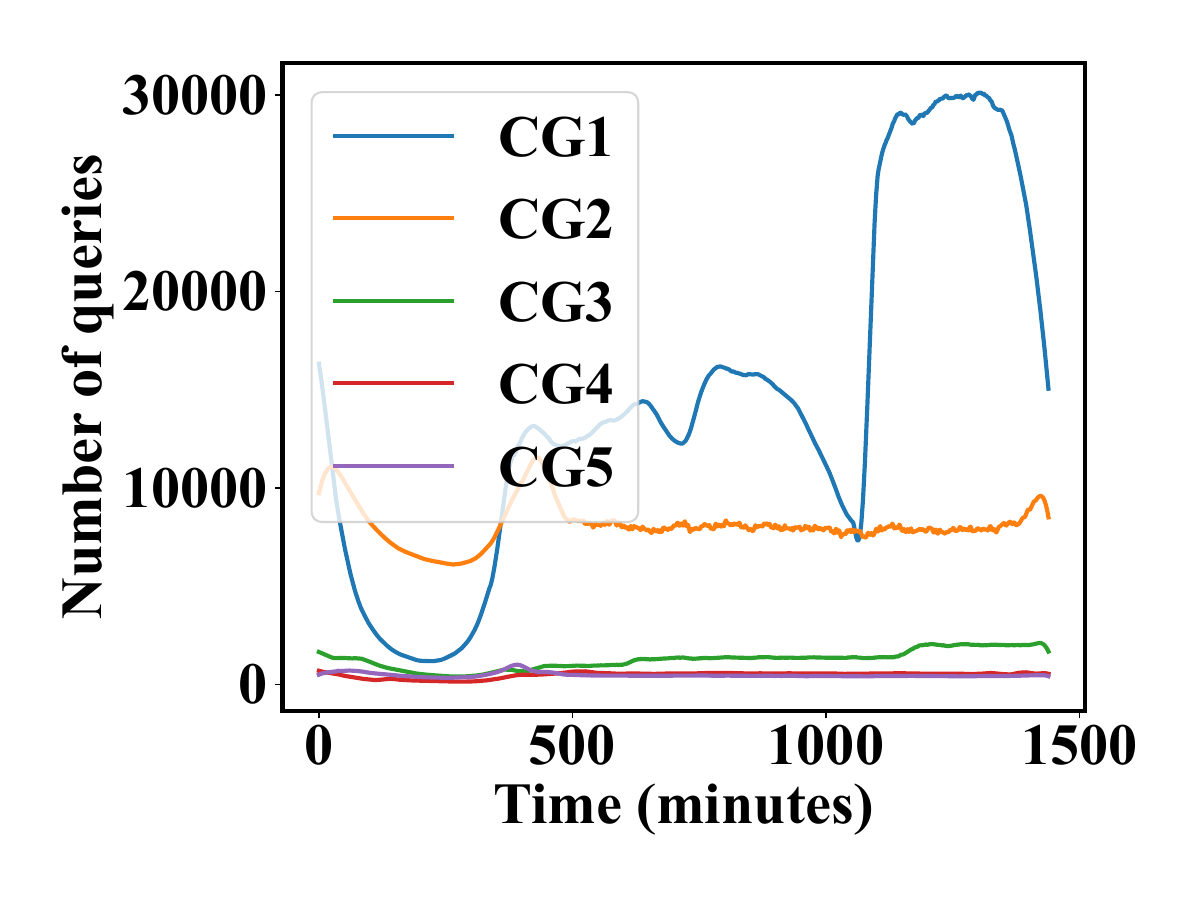}
        \label{cg_query}
    }
    \caption{Total number of queries and call graph queries over time for the service \textit{S\_130831269} in the Alibaba trace. }
    \label{dependency_dynamic}
    \vspace{-3mm}
\end{figure}

\subsubsection{Dynamic Service Dependency Graphs\label{section:problem_model}}
Service dependency graphs in production exhibit dynamic behaviors, with different call graphs changing their proportions over time. The occurrences of services in Alibaba traces follow a long-tailed distribution, where a small number of  service dependency graphs contribute to the majority of query counts. 
We remove the long tail and retain only the top 90\% of occurrences for the 1611 services.
The remaining 10\% of services have an average of only 8.5 requests per day, with 92.6\% receiving only one query per day. We believe that they are not suitable for feature analysis. A similar process is performed to retain the top 90\% of the call graphs, with a total number of 72,942. 
On average, there are more than 45 call graphs per dependency graph.

For a service, the number of queries per minute changes over time, as do queries accessing each call graph. For example, Fig.~\ref{dependency_dynamic} shows the total query number and the number of queries accessing each call graph over time, for the service with the highest number of queries (S\_130831269) in the dataset.
We can observe obvious variations in both the total queries and queries accessing different call graphs.

\subsubsection{Topological Differences in Dependency Graphs\label{section:dependency_differences}}
Different service dependency graphs in production exhibit significant differences in topology and invocation patterns. 

In terms of the topology, the depth and width of these graphs vary greatly. For the top 20 services with the most queries, the top-4 and 7 have a depth of 2 while top-8 and 11 reach a depth of 6. The width also varies, for example, the top-2 and 19 have a width less than 2, while the top-4 and 7 exceed 14.
In terms of invocation patterns, specific patterns are unique to certain dependency graphs. For instance, the self-invocation pattern (i.e., a microservice calls itself) is only present in top-1, 8, and 11. Moreover, unlike the other services, the top-1, 4, and 7 do not include calls to databases or memcached.

\subsubsection{Insights from Characterizing Dependency Graphs}
Production microservice traces contain vast amounts of data with highly dynamic and complex characteristics and often suffer from high missing rates~\cite{huye2023lifting}.
To utilize production microservice data for research, it is necessary to extract features to generate simulated service dependency graphs.
Whereas there are large topological and invocation pattern differences between service dependency graphs, we should categorize them using clustering methods to study features and construct the benchmark according to each category.

\subsection{Characterizing Call Graphs\label{trace_analysis_CGs}}
Both Alibaba and Meta traces are used in this subsection, and the key observations are as follows. 
1) Microservices overlap among the children sets of different call graphs.
2) Some microservices within a children set may be called multiple times. 
3) The probability of a microservice invoking different children sets is influenced by its sibling microservices. 


\subsubsection{Characteristics of Children Sets\label{section:children_set}}
\begin{figure}
    \centering
    \subfigure[Alibaba trace.]{
        \includegraphics[width=0.43\linewidth]{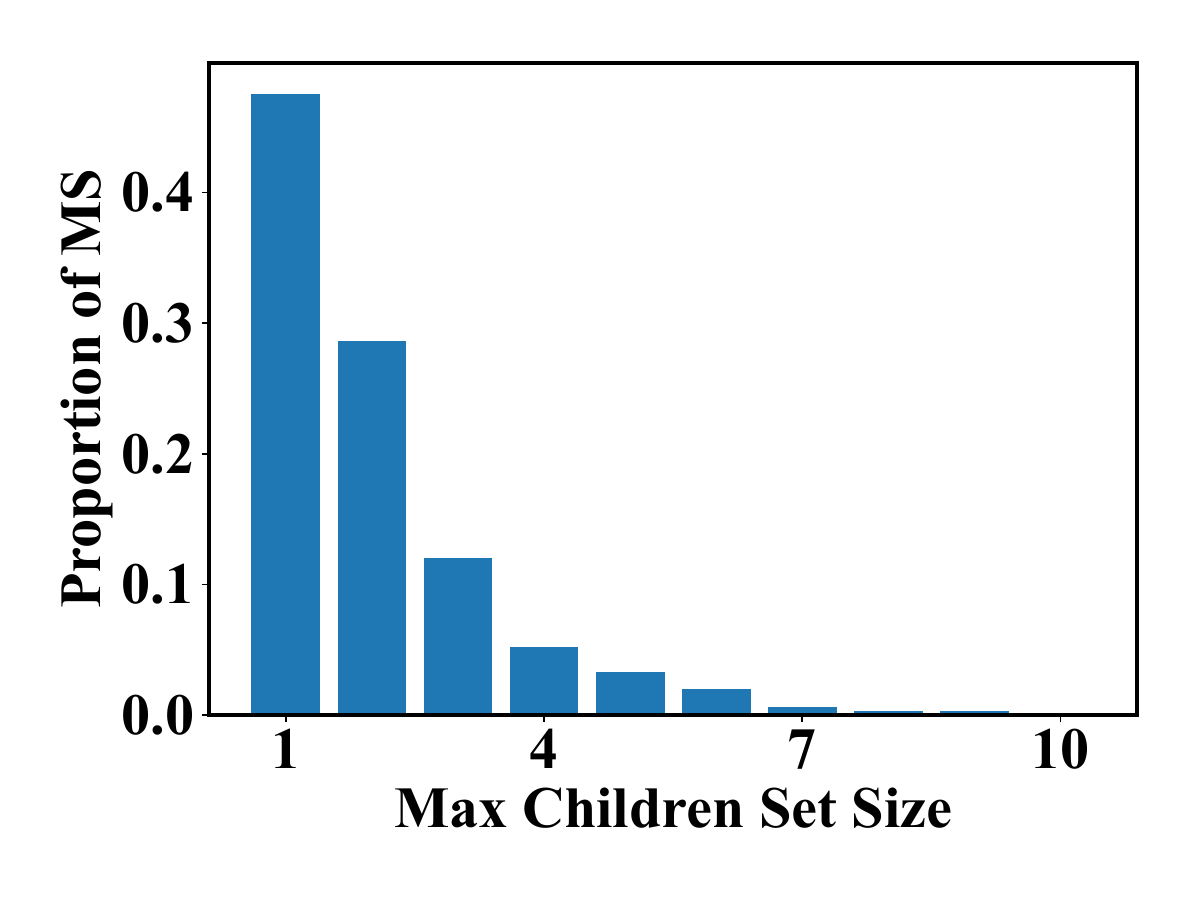}
        \label{total_query}
    }
    \hspace{5mm}
    \subfigure[Meta trace.]{
        \includegraphics[width=0.43\linewidth]{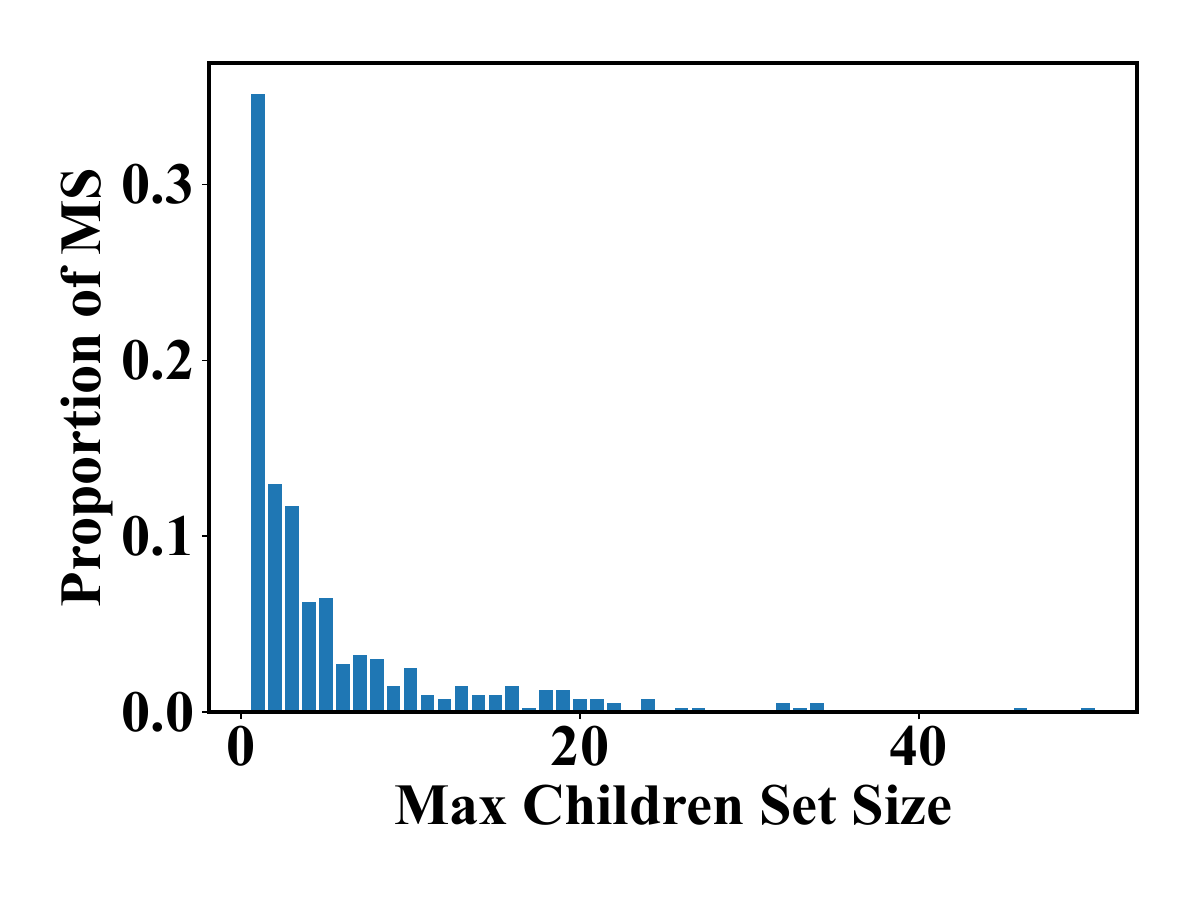}
        \label{cg_query}
    }
    \caption{Distribution of children set sizes for microservices.}
    \label{fig:children_set_size}
    \vspace{-3mm}
\end{figure}

Fig.~\ref{fig:children_set_size} shows the distribution of children set sizes in both traces. The children set size is the number of different microservices contained within a single children set. In the Alibaba trace, the sizes of the children sets range from 1 to 10, while those in Meta can be as large as 50. There are two major characteristics of microservice calls to children sets.

\textbf{Overlap in Children Sets:} Many microservices are shared by queries from different call graphs. There is a significant overlap in the children sets of different call graphs. In the Meta trace, 92.2\% of microservices appear in different children sets. Similarly, in the Alibaba trace, this overlap rate is 77.1\%.

\textbf{Repeated Calls:} There are widespread repeated calls from UM to DM. 
In the Meta trace, 20.4\% of repeated calls occur within children sets, and this value is 16.2\% for the Alibaba trace. 
Notably, database and memcached microservices exhibit a much higher rate of repeated calls at 52.5\% in the Alibaba trace. 
Tables~\ref{table:alibaba_meta_time} provide detailed statistics on the number of repeated calls in both datasets.



\begin{table}[ht]
\centering
\scriptsize
\caption{Number of repeated calls to each microservice in the traces}
\begin{tabular}{lccccc}
\toprule
 \textbf{Meta Trace} & \textbf{Min} & \textbf{Median} & \textbf{Mean} & \textbf{P99} & \textbf{Max} \\
\midrule
\textbf{Total} & 2 & 17 & 210 & 2,339 & 2,392 \\
\bottomrule
\end{tabular}
\\[3mm]
\begin{tabular}{lcccc}
\toprule
  \textbf{Alibaba Trace} & \textbf{Total} & \textbf{Database} & \textbf{Memcached} & \textbf{Other MS} \\
\midrule
\textbf{Min} & 2 & 2 & 2 & 2 \\
\textbf{Median} & 3 & 3 & 5 & 3 \\
\textbf{Mean} & 16 & 13 & 30 & 12 \\
\textbf{P99} & 374 & \textbf{198} & \textbf{469} & 75 \\
\textbf{Max} & 1080 & \textbf{397} & \textbf{537} & 1080 \\
\bottomrule
\end{tabular}
\label{table:alibaba_meta_time}
\vspace{-3mm}
\end{table}


\subsubsection{Sibling Set Influence\label{section:sibling_influence}}
Sibling microservices of a UM (i.e., the sibling set) can influence its probability of invoking DMs. 
Given a service dependency graph \(G(V, E)\) with microservices \(V\) and invocation relationships \(E\). \(G\) can include multiple call graphs \(CG_i(V_i, E_i)\), we can analyze the influence of sibling microservices. 
For each call graph \(CG_i\), microservice \(m\) has a sibling set \(S_i\) (nodes sharing the same UM as \(m\)) and a children set \(C_i\) (nodes invoked by \(m\)). 

\(P(u \to C_i)\) represents the likelihood that microservice \(u\) invokes the children set \(C_i\). It is calculated as the proportion of queries where \(u\) invokes \(C_i\) out of all queries where \(u\) invokes any possible children set.
\(P(u \to C_i \mid S_j)\) represents the likelihood that microservice \(u\) invokes the children set \(C_i\) given that the sibling set of \(u\) is \(S_j\). 
When there exist \(P(u \to C_i \mid S_j) \neq P(u \to C_i)\), we say that the probability of microservice \(u\) invoking the children set is influenced by the sibling set.
In extreme cases, we might have \(P(u \to C_i \mid S_j) = 1\), meaning the sibling set completely determines the children set that \(u\) invokes.

In the Alibaba trace, 55.7\% of the services are influenced by the sibling sets, where 92.7\%, 64.8\%, and 76.2\% of the microservices are influenced at call depths of 3, 4, and 5, respectively. This variation relates to the number of microservices at each depth: more microservices at the same depth lead to more sibling set combinations impacting children set calls.
The sibling set influence is determined by the nature of the microservice architecture. A user query triggers inter-microservice calls between related microservices, meaning those in the same call graph are functionally related. In some services, this functional correlation is more obvious, demonstrating sibling set influence.

\begin{figure}
  \centering
  \includegraphics[width=0.7\linewidth]{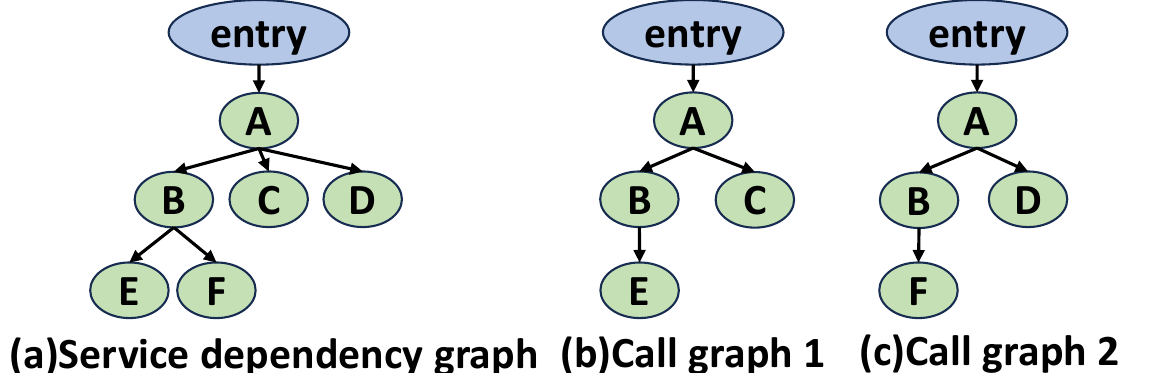}
  \caption{An example of a sibling set influence.}
  \label{fig:brother_influence_example}
  \vspace{-3mm}
\end{figure}

This finding creates opportunities for more accurate call graph generation and microservice architecture optimization.
First, modeling microservice calls as a probability model influenced by sibling sets better reflects the real-world production cluster scenarios, rather than considering only the two-level calls between UM and DM.
Second, microservices with significant sibling set influence can be reorganized into larger microservices. 
In Fig.~\ref{fig:brother_influence_example}, microservice $B$ calls $E$ when its sibling set is $\{C\}$ and calls $F$ when its sibling set is $\{D\}$. This suggests that $B$, $C$, and $E$ are tightly coupled and can be merged to reduce inter-microservice call overheads. Similarly, $B$, $D$, and $F$ can also be merged.

\subsubsection{Insights from Characterizing Call Graphs}


To generate more realistic simulated call graphs from production traces, we should model downstream microservices as children sets and consider richer call information, including repeated calls and sibling set influence.
Moreover, both fine-grained tracing and resource management of repeated microservice calls are essential for more effective microservice resource scaling.

\subsection{Characterizing Individual Microservices}
\label{section:dynamic_interface}
Since the Meta trace~\cite{metatrace} lacks detailed microservice interface information, we analyze individual microservices using the Alibaba trace~\cite{alitrace}.
We have two key observations:
1) Significant variation exists in the number of interfaces provided by microservices with different communication patterns.
2) Microservice interfaces are called by different call graphs.

\subsubsection{Types of Microservice Interfaces\label{section:interface_statistics}}
We observe that the microservices {\it memcached} and {\it http-called} have more interfaces, while other microservices are simpler and have fewer interfaces.
Most of the microservices in the Alibaba dataset are relatively simple; 88.31\% of them have less than 10 interfaces.


\begin{table}
    \centering
    \scriptsize
    \caption{Number of interfaces across communication modes}
    \label{tab:interface_counts}
    \begin{tabular}{lccccc}
        \toprule
        & \textbf{db} & \textbf{http} & \textbf{mc} & \textbf{mq} & \textbf{rpc} \\
        \midrule
        \textbf{Min} & 1 & 1 & 1 & 1 & 1 \\
        \textbf{Median} & 1 & 1 & 2 & 1 & 2 \\
        \textbf{Mean} & 2 & \textbf{11303} & 31 & 3.0 & 9 \\
        \textbf{P99} & 12 & \textbf{14731} & 26 & 39 & 74 \\
        \textbf{Max} & 22 & \textbf{1355296} & 2728 & 76 & 89 \\
        \bottomrule
    \end{tabular}
    \vspace{-3mm}
\end{table}
The number of microservice interfaces varies significantly across different communication modes. 
Table~\ref{tab:interface_counts} presents statistics on the number of microservice interfaces under various communication modes. 
The number of interfaces for HTTP-type microservices is notably higher.
We consulted the authors of the Alibaba trace regarding the HTTP interfaces.
They informed us that the number of interfaces in an HTTP microservice (e.g., entering microservice) is often correlated with the parameters involved, which can lead to a high interface count, particularly when multiple parameters are present.


\subsubsection{Interface Calling Patterns\label{section:interface_pattern}}
We observe that microservices may be shared by call graphs from different user queries, and these different call graphs may call different interfaces of the microservice. In the Alibaba trace, there are five main patterns of call graphs calling microservice interfaces.

a) Some microservices are only called by the same call graph, which calls the same interface of the microservice each time. 
b) Microservices called by the same call graph may also call different interfaces. 
c) Different call graphs call the same interface of the microservice. 
d) Multiple call graphs share the microservice and each call graph calls a different interface of the microservice. 
e) For a microservice, some call graphs call the same interface, and some call different interfaces.

\subsubsection{Insights from Characterizing Individual Microservices}
We should consider different interface calling patterns in the model of individual microservices when generating simulated service dependency graphs from production traces.
Moreover, it is beneficial to conduct fine-grained tracing and resource management of different microservice interfaces, to improve resource scaling efficiency.

\section{Dependency Graph Generation with DGG\label{section:benchmark_gen}}
In this section, we introduce the overview of DGG, followed by the design details and theoretical analysis.

\subsection{Overview of DGG}

\begin{figure}
  \centering
  \includegraphics[width=1\linewidth]{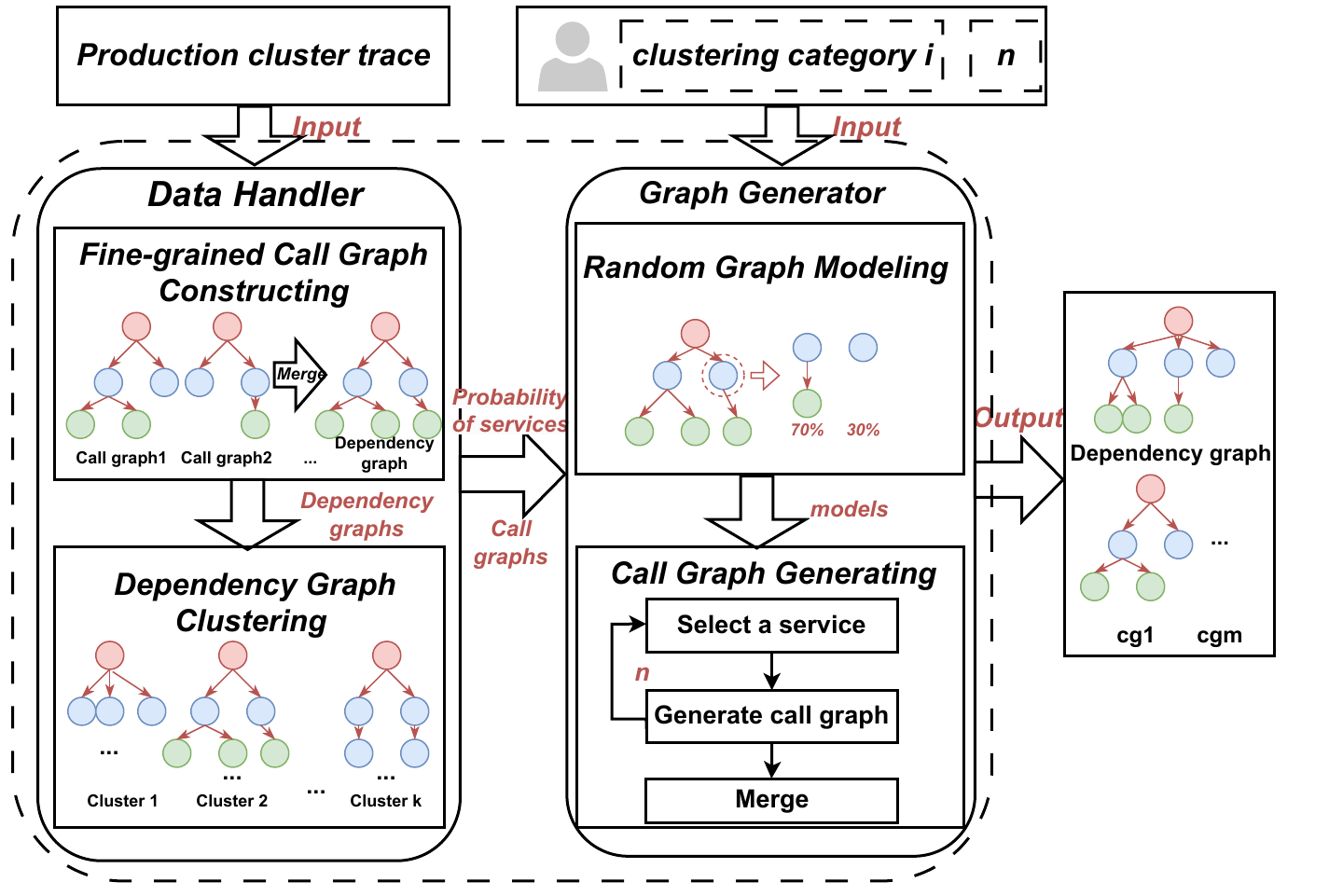}
  \caption{Design overview of DGG. 
  }
  \label{fig:benchmark_generate_overview}
 \vspace{-3mm}
\end{figure}

As shown in Fig.~\ref{fig:benchmark_generate_overview}, we design and implement a {\bf Service Dependency Graph Generator (DGG)} to generate microservice benchmarks that simulate realistic characteristics of services in production cluster traces. 
DGG consists of a \textit{Data Handler} and a \textit{Graph Generator}.
The data handler is responsible for organizing the large amount of data in the original production cluster trace.
Based on the organized data from the data handler, the graph generator takes the clustering category ($i$) and call graph generation time ($n$) from the user, adopting two steps to generate the service dependency graphs.

Based on the trace analysis observations, microservice calls often exhibit repeated invocations, dynamic interfaces, and significant differences among services. Therefore, DGG's data handler adopts two steps to address these characteristics. First, the \textit{fine-grained call graph constructing} constructs precise call graphs from the production traces and merges them into different dependency graphs. Then, the \textit{dependency graph clustering} classifies dependency graphs into different categories based on their topological and invocation characteristics.

To generate simulated call graphs that resemble realistic ones, DGG's graph generator first builds a graph probabilistic model for each service. Based on the observations in Section~\ref{trace_analysis_CGs}, it models the downstream microservices as children sets and comprehensively includes the impact of sibling sets of microservice calls, rather than modeling the relationships solely as a two-level structure between UM and DM. 
Using these models, the \textit{Call Graph Generating} then generates $n$ call graphs and merges them to the service dependency graph.



DGG works in the following steps. 
1) The data handler constructs precise call graphs from production traces and merges them into different dependency graphs. 
2) The data handler then clusters the dependency graphs into different categories based on their topological and invocation characteristics. 
It also obtains the probability of occurrence of each service in each category. 
3) With user inputs of the service category $i$ and the number $n$ of call graph generated, the graph generator builds a random graph model for each service in the category \(i\). 
4) Based on the random models, the graph generator then generates the call graphs $n$ times, with each time selecting a model of a service according to their probabilities of occurrence.
5) At last, the graph generator merges all call graphs to form the service dependency graph. 

DGG is implemented in Python, and we use 
$GraKeL$ library~\cite{JMLR:v21:18-370} to measure the similarity between dependency graphs. The statistical code lines of DGG are about 3000 lines.




\subsection{Data Handler}
In this subsection, we introduce the design of fine-grained call graph constructing and dependency graph clustering.

\subsubsection{Fine-grained Call Graph Constructing\label{FGCS}}
\begin{figure}
  \centering
  \includegraphics[width=0.85\linewidth]{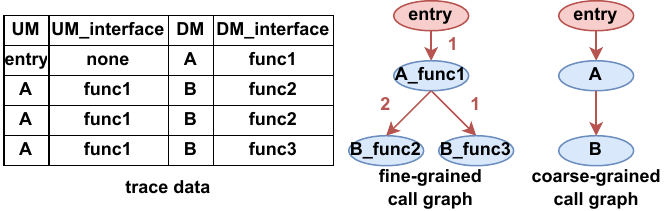}
  \caption{An comparison example of fine- and coarse-grained call graphs.}
  \label{fig:fine_grained_cg_example}
 \vspace{-3mm}
\end{figure}

Prior works~\cite{luo2021,luo2022erms} construct call graphs as directed acyclic graphs, with vertices representing microservices and edges representing invocations. However, this method fails to capture the repeated calls and dynamic interfaces common in production microservices.
To obtain fine-grained call graphs, we represent microservice invocations triggered by queries using weighted directed graphs.
We define vertices by combining the microservice name with the interface being called. 
The edges not only indicate invocation but also use weights to represent the number of times a DM is repeatedly called. 
Fig.~\ref{fig:fine_grained_cg_example} shows an example of fine-grained and coarse-grained call graphs generated from the same trace data.
For each service in the trace dataset, we construct all its fine-grained call graphs and merge them into the service dependency graph.

Since repeated calls and dynamic interfaces also impact microservice resource usage, this call graph constructor can be utilized for online microservice resource scaling to enhance resource allocation efficiency. We will explore its contributions to resource efficiency in Section~\ref{section:case_study_2}.

\subsubsection{Dependency Graph Clustering\label{section:cluster}}
To cluster the service dependency graphs by topological features and invocation properties, 
we use the graph kernel method~\cite{10.5555/3104322.3104356,JMLR:v21:18-370} to measure the similarity between dependency graphs, and then use K-means~\cite{electronics9081295,hamerly2003learning} to cluster all the dependency graphs. The inputs to the graph kernel include the adjacency matrix of dependency graphs, the microservice labels (e.g., db, memcached, normal), and the communication modes (e.g., http, rpc, mq), and the $K$ value representing the clustering category number.
The $K$ value is selected with the highest silhouette coefficient, which is a metric to measure the clustering quality. 
Each cluster groups services with similar topological and call characteristics, enabling users to choose specific clusters for generating simulated graphs.

Adopting the above clustering method on the Alibaba microservice trace~\cite{alitrace}, we categorize all the service dependency graphs into 6 classes.
We use the graph kernel method to quantify the similarity of service dependency graphs within each cluster and between different clusters. The intra-cluster similarity was found to be 3.4X higher than the inter-cluster similarity, proving our clustering accuracy.

\subsection{Graph Generator}
In this subsection, we introduce the process of building random graph models, and the algorithms for generating call graphs and dependency graphs using these models.

\subsubsection{Random Graph Modeling}


The variables used in the random graph modeling are shown in Table~\ref{table:variables}. 
For each service \( S \) in the trace dataset, we establish a random model \( G_s(V_s, E_s) \) based on its real dependency graph \( G(V, E) \).

The \( V \) represents the set of vertices in \( S \)'s service dependency graph, where each \( v \in V \) is a triplet representing the microservice's name, the interface being called, and its label, as \( (ms\_name, interface, label) \). The label set is
\(L = \{\text{database}, \text{memcached}, \text{normal}, \text{relay}, \text{leaf}\}\).
The label of a microservice can be determined by simple rules. If the communication mode is 'db' and 'mc', the labels are 'memcached' and 'database', respectively. For other communication modes, we examine whether the microservice continues to call other microservices. If it calls others, the label is 'relay'. If it never calls others, the label is 'leaf'. Otherwise, the label is 'normal'.

The \( E \) denotes the edge set of \( S \)'s dependency graph, where each edge is represented as a tuple \( (u, v, w, t) \) with \( u \) and \( v \) representing the source and destination microservices, \( w \) indicating the number of times \( v \) is repeatedly called, and \( t \in T = \{\text{http}, \text{rpc}, \text{mq}, \text{mc}, \text{db}\}\) representing the communication mode, which can be directly obtained from the trace. 

\begin{table}
\centering
\scriptsize
\caption{Variables used in Section 5.3}
\begin{tabular}{c|>{\centering\arraybackslash}m{6cm}}
\hline
\textbf{Variable} & \textbf{Description} \\ \hline
\(S\) & A real service in the dataset \\ \hline
\(V\) & Collection of all microservices in \(S\) with microservice names, interfaces, and labels \\ \hline
\(E\) & Invocation edges in \(S\) with UM, DM, number of invocations, and communication mode \\ \hline
\(L\) & Set of all possible microservice labels \\ \hline
\(T\) & Set of all possible communication modes \\ \hline
\(CG\) & Set of all call graphs of \(S\) \\ \hline
\(cg_i\) & The \(i\)-th call graph of \(S\) \\ \hline
\(u\) & A microservice in \(S\) \\ \hline
\(C_u\) & Set of all children sets of \(u\) in \(CG\) \\ \hline
\(C\) & A children set of \(u\) with UM, number of calls, and communication mode \\ \hline
\(d\) & Depth of \(u\)'s call in the call graph \\ \hline
\(s\) & Sibling set of \(u\) \\ \hline
\(\text{count}(cg_i)\) & Number of times \(cg_i\) is queried in \(S\) \\ \hline
\end{tabular}
\label{table:variables}
\vspace{-3mm}
\end{table}

We model the probability of each children set being called by a microservice. 
For a microservice \( u \) in \( S \), let \( C_u \) represent all possible children sets that \( u \) may call. 
Each children set \( C \in C_u \) consists of vertices $v$ and the corresponding call edges $e$. 
Therefore, each children set includes the microservice name, interface, label, the number of repeated calls, and the communication mode.
An empty children set indicates \( u \) does not call any other microservices. 
Let \(CG\) denote the set of all call graphs for \(S\), where each call graph \(cg_i \in CG\) has an associated occurrence count \(count(cg_i)\).

Considering the sibling set influence, and the impact of microservices' call depth on children set calls, we characterize this influence using conditional probabilities.
Define the indicator variable \( I(u \to C , s, d, cg_i) \), when \( I(u \to C , s, d, cg_i) = 1 \) denotes that vertex \(u\) calls children set \(C\) at depth \(d\) with sibling set \(s\) in call graph \(cg_i\). The probability that \(u\) calls children set \(C\) given \(u\)'s sibling set \(s\) and depth \(d\) can be calculated in Equation~\ref{equation:pr2}.
\begin{equation}
\scriptsize
P(u \to C \mid s, d) = \frac{\sum_{cg_i \in CG} I(u \to C , s, d, cg_i) \cdot count(cg_i)}{\sum_{C' \in C_u} \sum_{cg_j \in CG} I(u \to C', s, d, cg_j) \cdot count(cg_j)}
\label{equation:pr2}
\end{equation}

The time to establish a random graph model for a real service depends on the service dependency graph scale. In our experiments, the average model establishment time is 4ms.

\subsubsection{Call Graph Generation Algorithm\label{sec:cg_gen}}
Based on the random graph model, the call graph generation process is as follows.
1) Initialization: The graph starts with the vertex \textit{entry} as the beginning of the query.
2) Vertex and Edge Addition: For each vertex \(u\) labeled as \textit{normal} or \textit{relay}, determine its sibling set \(s\) and depth \(d\). Based on Equation~\ref{equation:pr2}, choose a children set \(C\) and add it to the graph. For each \((v, w, t)\) in \(C\), add edge \((u, v, w, t)\) to the graph.
3) Iterative Expansion: Repeat for each new vertex \(u\) until no further vertices can be added.

Algorithm~\ref{alg:call-graph-generator} shows the major process of generating a call graph. The output is a call graph \(G\) stored as a list of call relationships.
Each element in the list has the structure $(um\_depth, UM, DM, weight, compara)$, where $um\_depth$ represents the depth of the $UM$ in the call graph, $UM$ and $DM$ denote the respective vertices, $weight$ indicates the number of times $DM$ is repeatedly called, and $compara$ signifies the communication mode.
A queue \(Q\) stores vertices that are not handled. Initially, \(G\) is empty, and \((1, (entry, none, relay))\) is pushed into \(Q\) (line \ref{alg:init_push}).

In the main loop, the algorithm pops a vertex and its depth \((d, UM)\) from \(Q\) (line \ref{alg:pop}). If \(UM\) is labeled \textit{relay} or \textit{normal}, it determines the children set based on the random graph model (lines~\ref{alg:get_sibling}-\ref{alg:consider_sibling}). 
For other labels, the algorithm skips to the next iteration. For each target vertex in the children set, it generates a new call relationship edge, adds it to \(G\), and pushes the new vertex into \(Q\) (lines \ref{alg:for_target}-\ref{alg:push}). The process continues until \(Q\) is empty, then returns the call graph \(G\).


\begin{algorithm}
\caption{Call Graph Generator}
\scriptsize
\label{alg:call-graph-generator}
\begin{algorithmic}[1]
\ENSURE $G$: A call graph stored as a list of call relations 
\STATE $Q \leftarrow$ queue to temporarily store vertices without generated $DM$
\STATE $G$ is initialized as an empty list
\STATE Push $(1, (entry,none,relay))$ into $Q$ \label{alg:init_push}

\WHILE{$Q$ is not empty}
    \STATE $(d, UM) \leftarrow Q.\text{pop}()$ \label{alg:pop}
    \IF{$UM.label$ == $relay$ or $normal$} \label{alg:haschild}
        \STATE $s \leftarrow \text{get\_sibling\_set}(UM)$\label{alg:get_sibling}
        \STATE $targets \leftarrow \text{children\_set}(UM, d, s)$\label{alg:consider_sibling}

        \FOR{$\text{target} \in \text{targets}$} \label{alg:for_target}
            \STATE $(DM, weight, t) \leftarrow (\text{target.v}, \text{target.w}, \text{target.t})$
            \STATE $G.\text{add}(d, UM, DM, weight, t)$
            \STATE $Q.\text{push}(d+1, DM)$ \label{alg:push}
        \ENDFOR
    \ELSE
        \STATE \textbf{continue} \label{alg:continue}
    \ENDIF
\ENDWHILE

\RETURN $G$
\end{algorithmic}
\end{algorithm}

Algorithm \ref{alg:call-graph-generator} has a time complexity of \(O(|V| \cdot k)\), where \(|V|\) is the total number of vertices in the call graph, and \(k\) is average size of the children sets. It uses an list to store edges of the generated graph, resulting in a space complexity of \(O(|E|)\), where \(|E|\) is the total number of edges in the call graph. In our experiments, the average time to generate a call graph is 0.7 ms.
We use actual microservice names from the dataset, effectively managing scenarios with multiple parents, depths, or shared microservices across call graphs.

After all call graphs are generated with the user-specified $n$ times, DGG merges them to form a service dependency graph.

\subsection{Theoretical analysis}
In this subsection, we analyze the probability distribution similarity in topology between call graphs generated by DGG and those observed in the real-world trace dataset.


We first show that for any given real service \( S \), the topological distribution of each real call graph within \( S \) is similar to the call graphs generated using \( S \)'s random graph model.
Then, using the law of total probability, we establish that the topological distribution of all real services is similar to that of all generated call graphs. We examine three aspects: the width (microservices number) at each layer, the depth of the call graph (total layers), and the total number of microservices. 

The proof order of the three aspects follows the considerations below.
First, a call graph has exactly \( h \) layers of depth if and only if the size of the children set of all microservices at depth \( h \) is 0. This is related to the width of each layer.
Second, the total microservice number in a call graph is the sum of microservices at each layer. 
Thus, we choose to first prove the aspect of the width, then the depth, and finally the total microservice number.
The propositions are as follows.

\begin{proposition}
\label{propostion:width}
Let \( M_k \) represent the number of vertices at the \( k \)-th level of the call graph. We denote \( P_{\text{gen\_width}}(M_k = m_k) \) as the probability that the \( k \)-th level of the call graph has \( m_k \) vertices according to a random graph model, and \( P_{\text{real\_width}}(M_k = m_k) \) as the probability that the \( k \)-th level of the real call graph has \( m_k \) vertices. Then,
\begin{equation}
\scriptsize
P_{\text{gen\_width}}(M_k =m_k) = P_{\text{real\_width}}(M_k =m_k)
\end{equation}
\end{proposition}
\begin{proposition}
\label{proposition:depth}
Let $H$ denote the total depth of a call graph.
Let $P_{\text{gen\_depth}}(H=h)$ denote the probability that a generated call graph has depth $h$, and $P_{\text{real\_depth}}(H=h)$ denote the probability of depth $h$ in the real dataset. Then,
\begin{equation}
\scriptsize
P_{\text{gen\_depth}}(H=h) = P_{\text{real\_depth}}(H=h)
\end{equation}
\end{proposition}
\begin{proposition}
 \label{proposition:num}
     Let $N$ denote the total number of vertices in a call graph, \( P_{\text{gen\_num}}(N=n) \) denote the probability that the generated call graph has \( n \) microservices. And $P_{\text{real\_num}}(N=n)$ denote the probability that the real call graph has \( n \) microservices. Then,
 \begin{equation}
 \scriptsize
     P_{\text{gen\_num}}(N=n) = P_{\text{real\_num}}(N=n)
 \end{equation}
\end{proposition}

The detailed proof can be found in \url{https://anonymous.4open.science/r/Theoretical-Proof-40B1}, if you are interested.

\section{Case Studies\label{sec:case_studies}}

In this section, we first utilize DGG to generate service dependency graphs and evaluate their similarity with real-world ones.
Then, we investigate another effect of DGG's call graph constructing (Section~\ref{FGCS}), i.e., promoting microservice resource scaling efficiency under dynamic loads online.

\subsection{Dependency Graph Similarity of DGG\label{sec:eval-similarity}}

\subsubsection{Investigation Setup}
Since the Meta trace~\cite{metatrace} lacks detailed calling information among microservices, 
the Alibaba trace~\cite{alitrace} is adopted in this section.
We compare DGG to a state-of-the-art microservice call graph generator (named CGG for short)~\cite{luo2021}. CGG computes the width of each layer and assigns labels to each microservice based on the probability distribution of the overall trace to generate a call graph. 
According to the six clustering types determined by DGG (Section~\ref{section:cluster}), we construct six service dependency graphs with each of 1,000,000 call graph generations. 
We compare all generated call graphs of DGG and CGG to real-world ones in
the graph topology (width, depth, and microservice number), microservice types, and communication modes.

We quantify the similarity by using Jensen-Shannon (JS) divergence, which is proved to be effective in comparing two probability distributions~\cite{briet2009properties}. Smaller JS divergence indicates closer distributions.
It is more comprehensive to compare each call graph's similarity than the direct service dependency graphs, as call graphs represent each query execution path 
which can capture dynamic interactions among microservices. 

\begin{figure}
    \centering
    \subfigure[Call graph percentages.]{
        \includegraphics[width=0.45\linewidth]{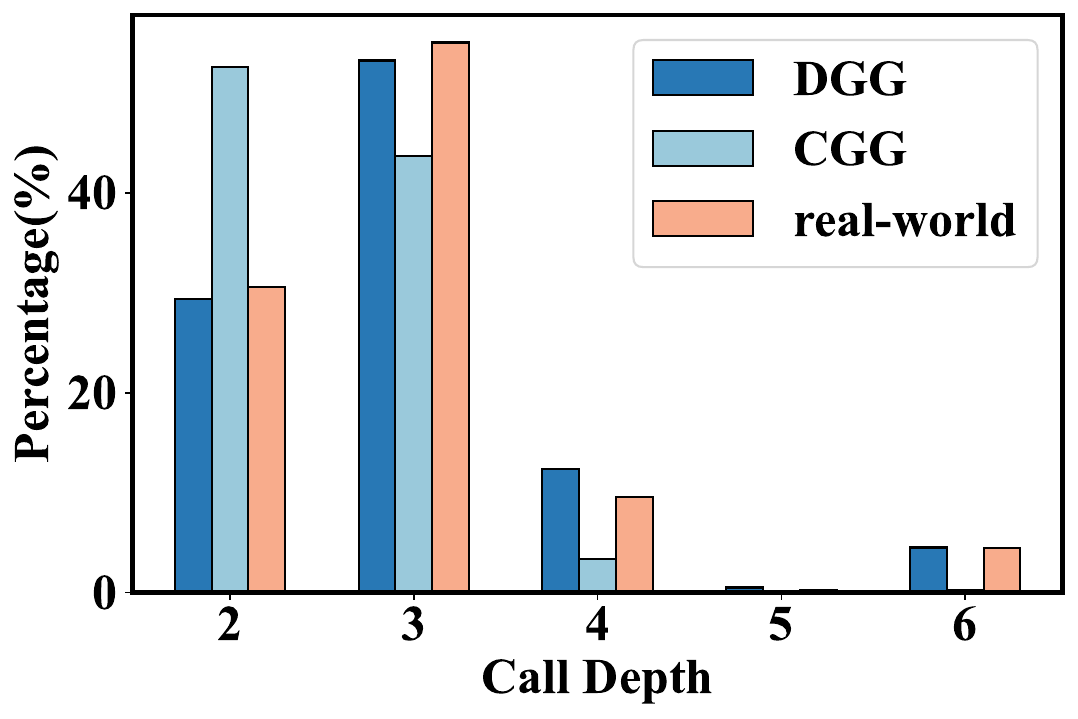}
        \label{fig:depth_comparison}
    }
    \subfigure[Expected microservice number.]{
        \includegraphics[width=0.45\linewidth]{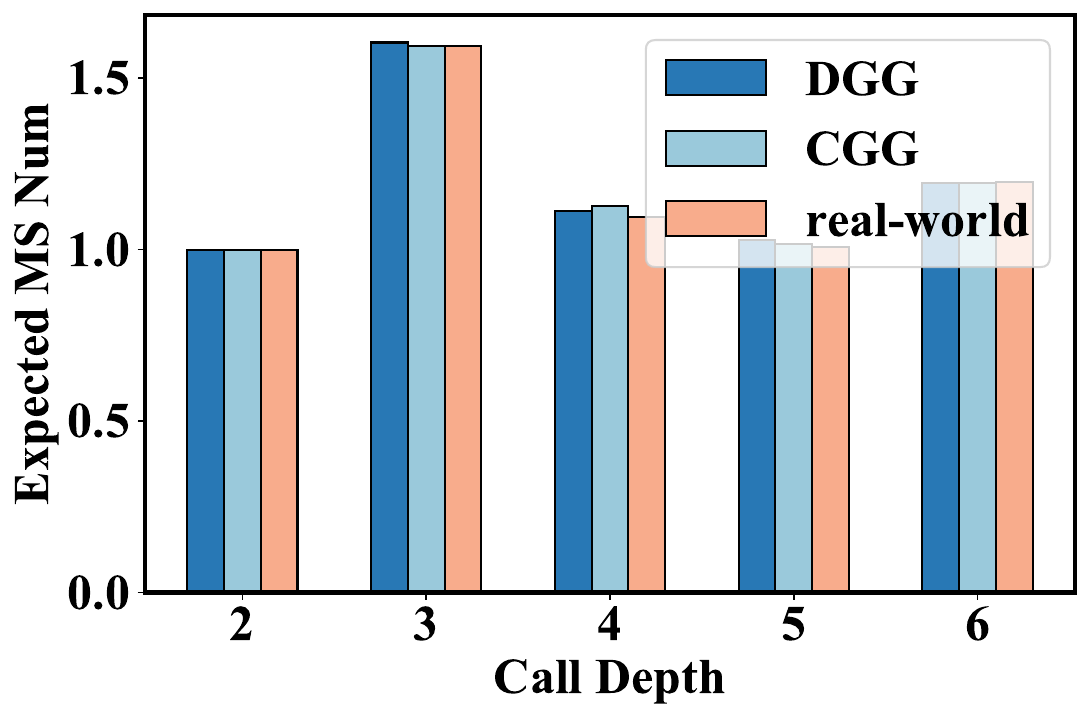}
        \label{fig:width_comparison}
    }
    \caption{Call graph percentages and expected microservice number under different depths of the DGG, CGG, and real-world dataset.}
    \label{fig:case_example}
    \vspace{-3mm}
\end{figure}

\subsubsection{Graph Topology Similarity}
Since the percentage of call graphs in depths greater than 6 is less than \(10^{-5}\) in the Alibaba trace, we compare call graphs with depths less than or equal to 6.
Moreover, the percentage of call graphs with microservice numbers greater than 14 is less than \(10^{-6}\), so we also focus on call graphs with numbers less than 14 in this subsection. 

Fig.~\ref{fig:depth_comparison} shows the call graph percentages under different depths of the DGG, CGG, and the real-world dataset, respectively.
DGG's call graphs closely match the distribution of real-world call graphs. By contrast, call graphs generated by CGG show a higher percentage at depth 2 and lower percentages for depths greater than 2. This indicates CGG has an early termination of calls in the generated call graphs. 
The reason could be that CGG has a higher probability of generating microservices labeled ``leaf'', ``memcached'' and ``db'' that do not continue to invoke DMs.
As statistics, the JS divergence of call graph percentage distributions between the DGG and real-world dataset is 0.034, whereas the value for CGG is 0.193. 

Fig.~\ref{fig:width_comparison} shows the expected number of microservices (i.e., layer width) under different depths. 
DGG and CGG are both relatively close to real-world dataset.
DGG's distribution is a little closer to real-world call graphs compared to CGG, with a JS divergence of 0.003 for DGG and 0.004 for CGG. 

\begin{figure}
  \centering
  \includegraphics[width=0.7\linewidth]{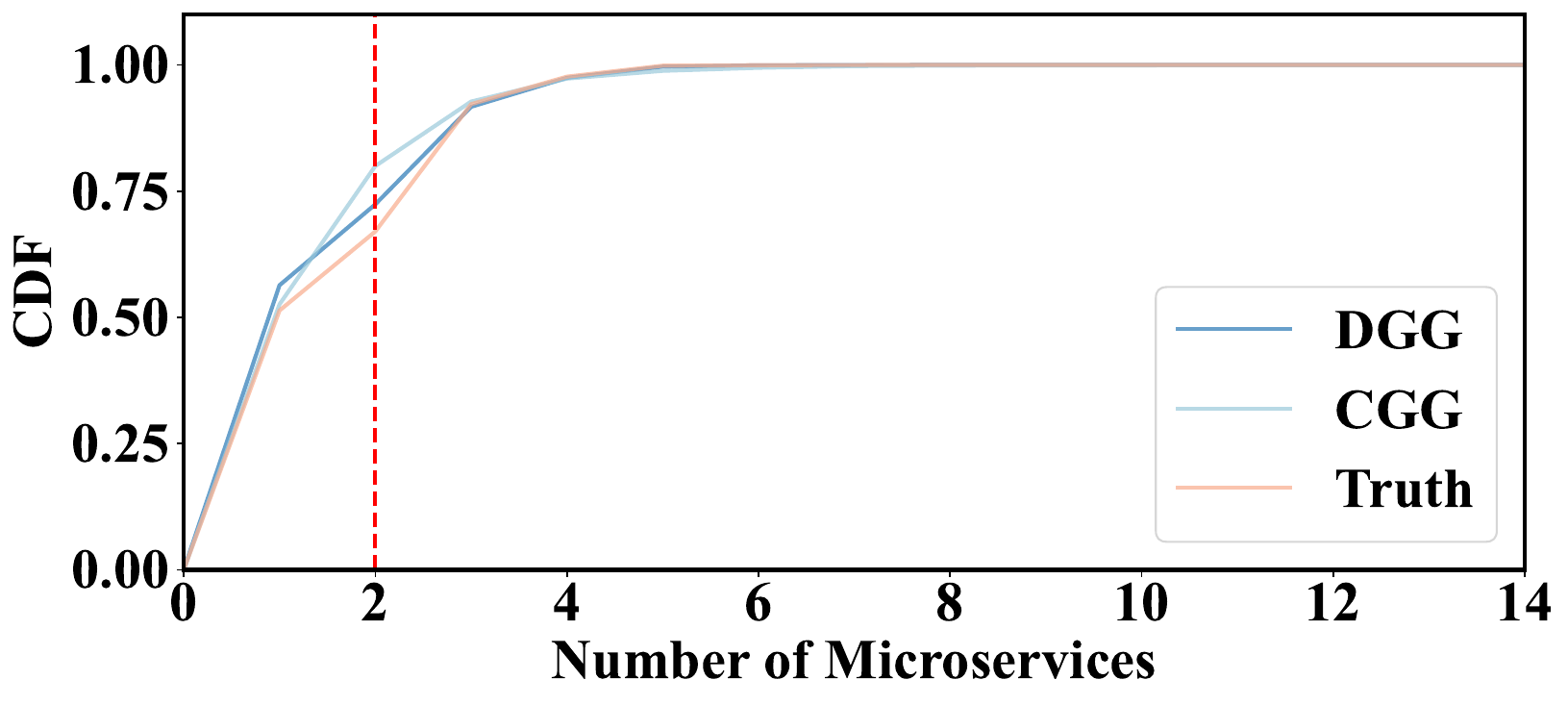}
  \caption{Cumulative call graph percentage distribution of the microsevice number. The red dotted line highlights that, for the same number of microservices, the DGG curve is closer to the Truth curve than the CGG curve.
  }
  \label{fig:size_comparison}
  \vspace{-3mm}
\end{figure}

Fig.~\ref{fig:size_comparison} shows the cumulative call graph percentage distribution of microservice number of DGG, CGG, and real-world dataset, respectively.
We can observe that DGG's curve closely matches the curve of real-world, while CGG's generated call graphs have fewer microservices. The JS divergence between DGG and real-world call graph percentage distributions is 0.053, while the value for CGG is 0.146.
\begin{figure}
  \centering
  \includegraphics[width=1\linewidth]{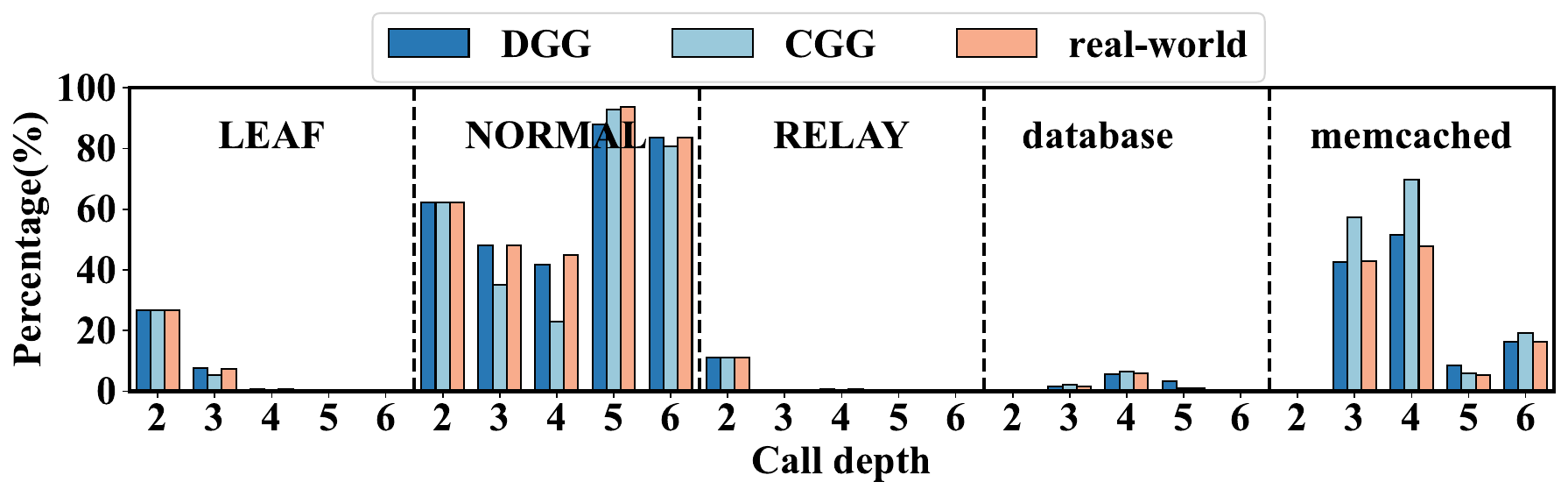}
  \caption{The percentages of different types of microservices under different depths of call graphs of DGG, CGG, and real-world dataset, respectively.
  }
  \label{fig:labels_comparison}
  \vspace{-3mm}
\end{figure}

\subsubsection{Microservice and Communication Similarity}
Fig.~\ref{fig:labels_comparison} shows the percentages of different types of microservices under different depths of call graphs. For each microservice type, the distribution in call graphs generated by DGG matches the real-world call graphs more closely. By contrast, call graphs generated by CGG show differences: the percentages of ``relay'' and ``normal'' labels are lower than in real call graphs, while ``memcached'', ``database'', and ``leaf'' labels are excessive. This difference indicates an early termination of calls in CGG generated graphs, as ``relay'' and ``normal'' may continue to call other microservices while ``memcached'', ``database'', and ``leaf'' terminate calls earlier.
\begin{figure}[t]
  \centering
  \includegraphics[width=1\linewidth]{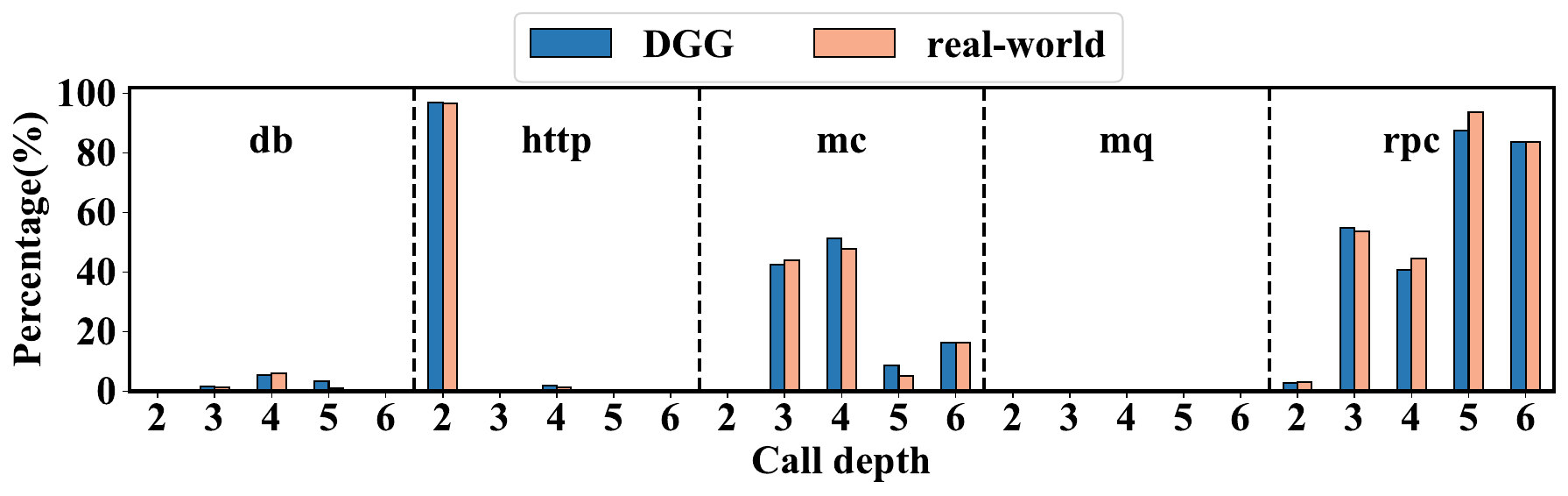}
  \caption{
  The percentages of callings to downstream microserivces in different communication modes under different depths of call graphs of DGG and real-world dataset, respectively.
  }
  \label{fig:comparas_comparison}
  \vspace{-3mm}
\end{figure}

Since CGG does not specify how communication modes between microservices are determined, we compare the similarity between DGG and real-world dataset in communication modes.
Fig.~\ref{fig:comparas_comparison} shows percentages of callings to downstream microservices in different communication modes under different depths of call graphs.
We can observe that DGG's generated call graphs closely approximate the distribution of communication patterns in the real-world.

In summary, the call graphs generated by DGG have more similar distributions in microservice types and communication modes than CGG to real-world dataset.

\subsection{Resource Efficiency with DGG’s Call Graph Constructor\label{section:case_study_2}}
DGG generates service dependency graphs that contain realistic characteristics for producing benchmarks to assist in resource management studies, rather than directly managing microservice resources.
Moreover, its call graph constructing (Section~\ref{FGCS}) can be integrated in typical microservice resource managers to enable more efficient resource scaling, as it can capture production characteristics of repeated calls and dynamic calling interfaces.
Therefore, we investigate the effectiveness of DGG's fine-grained constructing at online.
\subsubsection{Scaling Strategy and Baselines}
We integrate DGG's call graph constructing into resource scaling to form \textit{FineGrained-Scale}. First, we offline profile the computing resource demand (CPU cores) for each call graph to get the fine-grained resource demand of each microservice under different loads (from 3000 queries per minute to the maximum supported load of our cluster). Based on this profiling, we build offline linear regression models for each microservice in each call graph with input of load and output of resource demand, which is proven to be effective in state-of-the-art works~\cite{nodens,erms}.
At online, we use Jaeger~\cite{jaeger} to trace fine-grained call graph loads at one-minute interval, then input to corresponding models to determine the resource allocation for each microservice.

Prior microservice resource management works~\cite{firm,10.1145/3631607,erms,nodens,10.1145/3542929.3563477,suresh2017distributed} constructed coarse-grained call graphs, neglecting various interfaces and repeated callings, thus different fine-grained call graphs are mapped to the same coarse-grained call graph.
With coarse-grained call graphs, the aggressive~\cite{erms,firm,10.1145/3631607,suresh2017distributed} or conservative~\cite{nodens,10.1145/3542929.3563477} strategies can only be utilized for resource scaling, which scale according to the maximum or minimum possible computing resource usage (named MAX-Scale and MIN-Scale), respectively.
We adopt the above two strategies as baselines in our evaluations.
For MAX-Scale/Min-Scale, the computing resource of a microservice is allocated based on its coarse-grained total load and the linear model of FineGrained-Scale that has the maximum/minmum resource allocation value online.

\subsubsection{Investigation Setup}
\begin{table}
\centering
\caption{Experiment specifications}
\label{table:specifications}
\begin{tabular}{ >{\centering\arraybackslash}m{0.16\linewidth} | >{\centering\arraybackslash}m{0.75\linewidth} }
\hline
\textbf{} & \textbf{Specifications} \\ \hline
\textbf{Hardware} & Three-node cluster, Intel(R) Xeon(R) CPU E5-2630 v4 @ 2.20GHz, 128GB Memory Capacity, 25 MiB L3 Cache Size (20-way set associative) \\ \hline
\textbf{Software} & Ubuntu 20.04.6 LTS with kernel 5.15.0-107-generic, Docker version 24.0.5, Kubernetes version v1.20.4, Golang version 1.19.3, gRPC version 1.29.1 \\ \hline
\end{tabular}
\vspace{-3mm}
\end{table}

We use three sets of benchmarks.
First is DGG's generated dependency graphs (Section~\ref{sec:eval-similarity}), based on which we generate six \textit{Simulated benchmarks}.
Second is to select six dependency graphs from the remaining 20\% services in Alibaba traces as stated in Section~\ref{trace_overview}, and generate six \textit{Real-world benchmarks}
Third is the DeathStarBench's SocialNetwork (\textit{SN}). To better present real-world application characteristics, we enhance its compose-post service with four new microservices and six new call graphs to support various methods of text filter and image compression.

To implement \textit{Simulated} and \textit{Real-world benchmarks}, we use Memcached~\cite{memcached} and MongoDB~\cite{mongodb} for microservices labeled ``memcached'' and ``db'', respectively.
We implement stateless microservices in Golang~\cite{golang} and use commonly-used QuickSort, PageRank, and Word Stemming as their workloads, similar to previous works~\cite{mirhosseini2021parslo,mirhosseini2020q,nodens}.
Moreover, we adopt the Google gRPC~\cite{grpc} for RPC calling and Inter-Process Communication~\cite{luo2021} for calling stateful microservices.

In terms of call graph loads, for each simulated benchmark with n call graphs, we utilize the query per minute of the top-n accessed call graphs in the corresponding service clustering set in the trace.
Moreover, for each real-world benchmark, we directly utilize its realistic query per minute in the trace.
For \textit{SN}'s six call graphs, we utilize the query per minute of the top-6 call graphs with the most number of queries in the trace.
We evaluate each benchmark for one hour with the trace data and set the QoS of each benchmark to its 95\%-ile latency under no computing resource constraints. Table~\ref{table:specifications} shows the hardware and software configurations.

\subsubsection{Tail Latency and Resource Allocation}

\begin{figure}
  \centering
  \includegraphics[width=.9\linewidth]{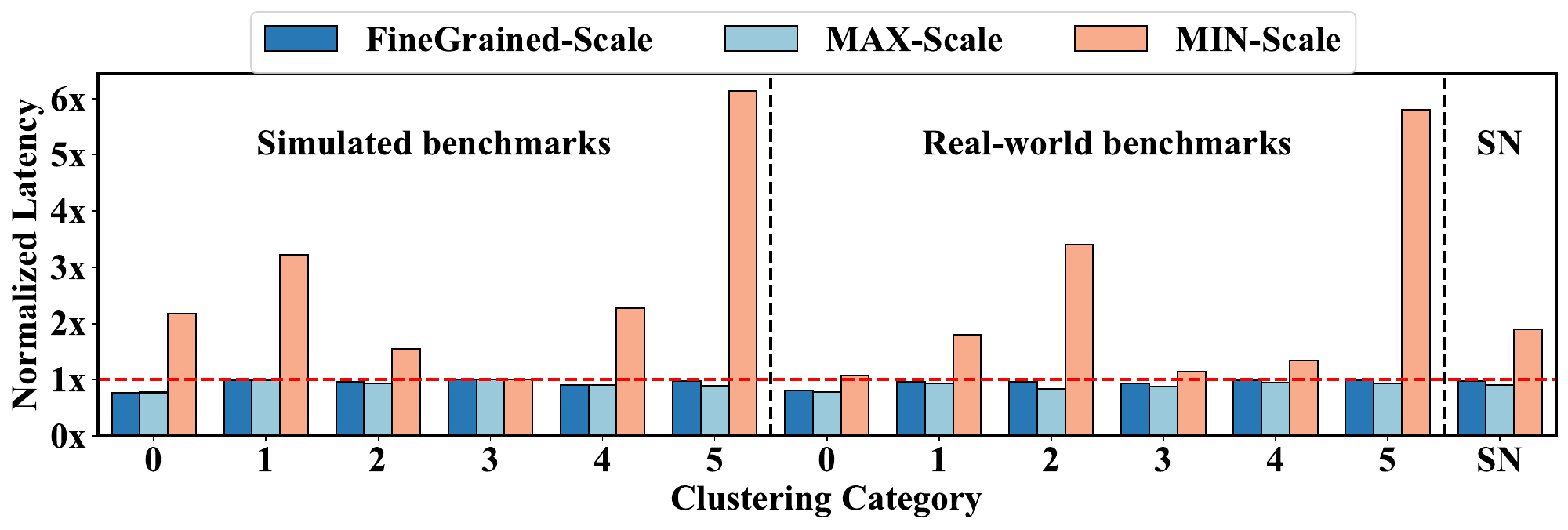}
  \caption{The 95\%-ile latencies of FineGrained-Scale, MAX-Scale, and MIN-Scale normalized to the QoS target, respectively.}
  \label{fig:normalized_latency}
  \vspace{-3mm}
\end{figure}

\begin{figure}
  \centering
  \includegraphics[width=.9\linewidth]{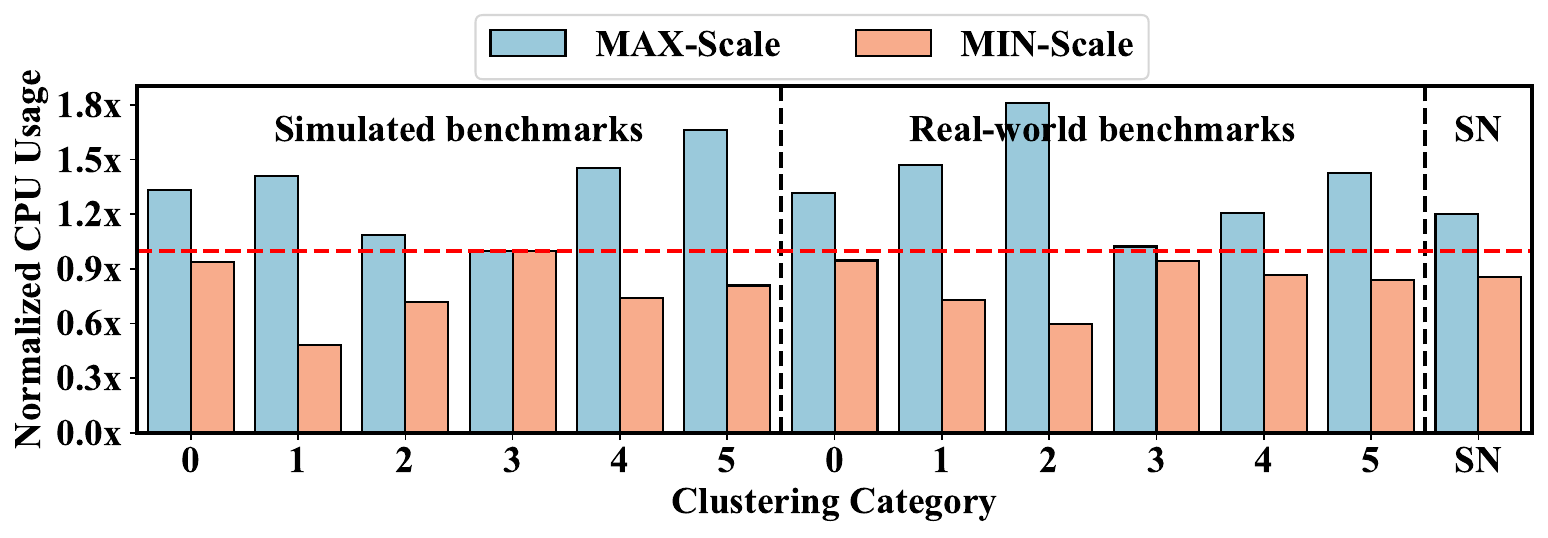}
  \caption{The CPU core hour usage in the evaluations of MAX-Scale and MIN-Scale normalized to FineGrained-Scale.}
  \label{fig:normalized_cpu} 
 \vspace{-3mm}
\end{figure}

Fig.~\ref{fig:normalized_latency} shows the 95\%-ile latencies of the benchmarks with three strategies. 
We can observe that both FineGrained-Scale and MAX-Scale can guarantee the QoS, while Min-Scale violates the QoS by 2.5X on average.
Fig.~\ref{fig:normalized_cpu} shows the total CPU core hour usage for all microservices with three strategies under different benchmarks. 
We can observe that FineGrained-Scale has less resource usage than MAX-Scale, with an average reduction of 25.3\% and a maximum reduction of 44.8\%. 
MIN-Scale has the lowest resource usage, but it has serious QoS violations.

FineGrained-Scale is able to fine-grained identify various interface callings and repeated callings to microservices, which can allocate just-enough computing resources while ensuring the QoS.  
By contrast, MIN-Scale and MAX-Scale can only construct coarse-grained call graphs. 
MIN-Scale allocates resources based on the microservice interface with the minimum resource demand and the single calling in each query and thus results in QoS violations.
MAX-Scale conservatively allocates resources based on the maximum possible demand for each microservice, leading to excessive resource allocation.

\subsubsection{Diving into High Scaling Efficiency}
\begin{figure}
    \centering
    \subfigure[The CPU core hour usage.]{
        \includegraphics[width=0.45\linewidth]{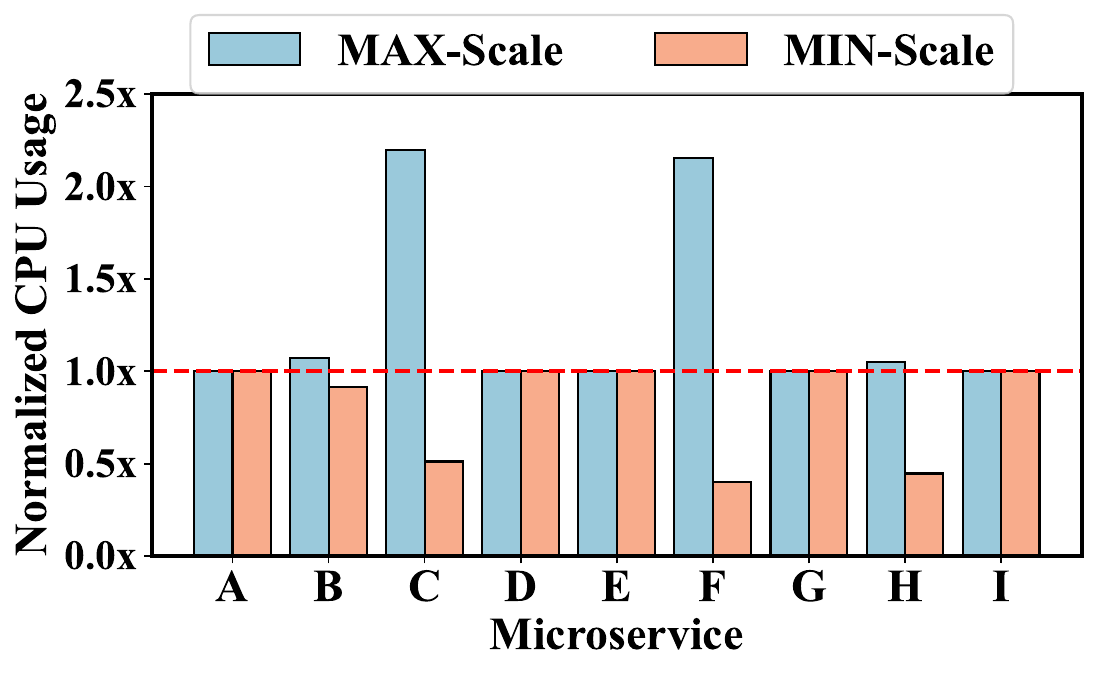}
        \label{fig:example_cpu}
    }
    \subfigure[The 95\%-ile latencies.]{
        \includegraphics[width=0.45\linewidth]{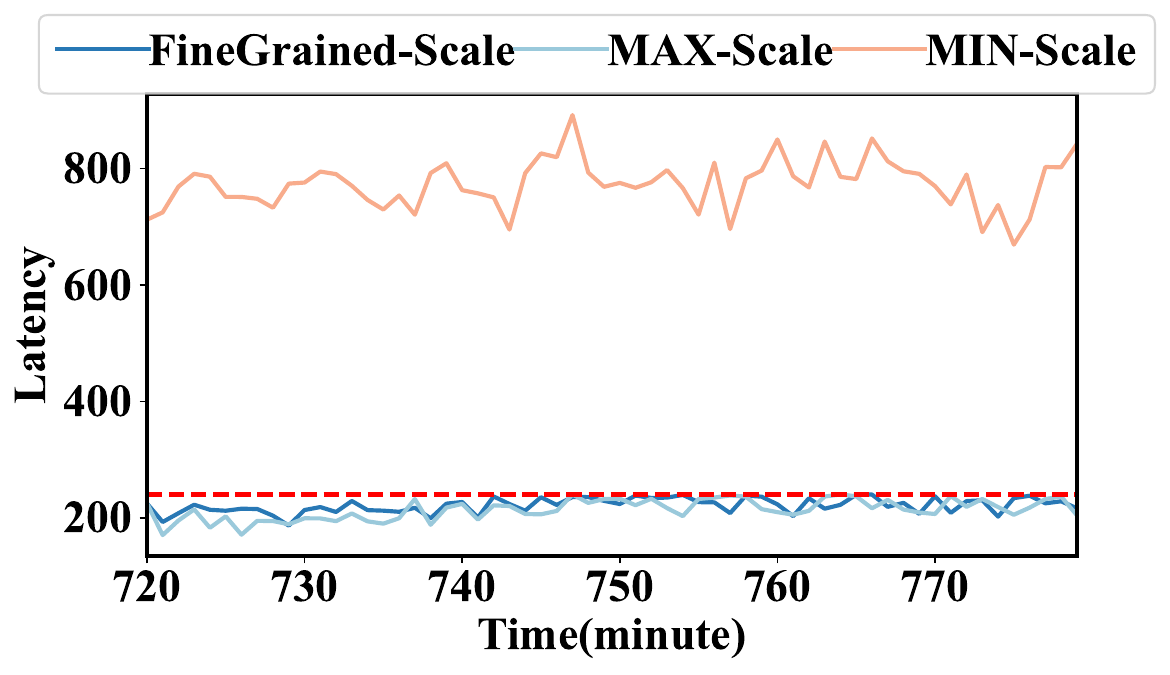}
        \label{fig:example_latency}
    }
    \caption{The CPU core hour for each microservice and 95\%-ile latencies under FineGrained-Scale, MAX-Scale, and MIN-Scale of an example benchmark.}
    \label{fig:case_example}
\end{figure}

\begin{figure}
  \centering
  \includegraphics[width=0.8\linewidth]{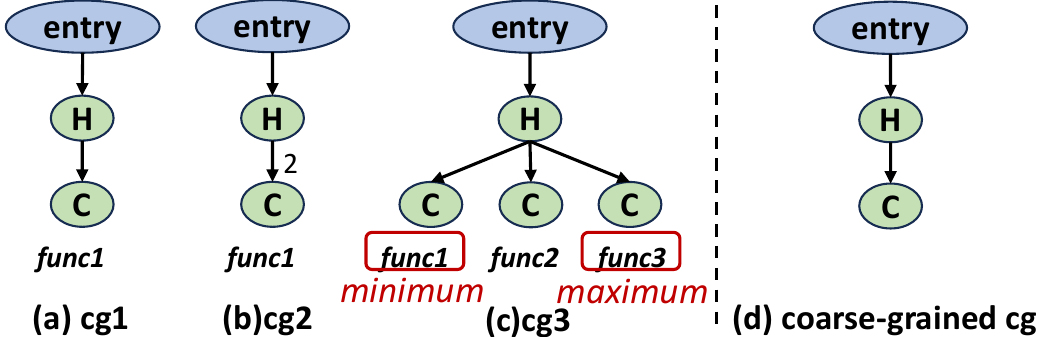}
  \caption{The fine- and coarse-grained call graphs of the example benchmark.}
  \label{fig:cg_example}
  \vspace{-3mm}
\end{figure}

We use an example benchmark to better understand the benefits of FineGrained-Scale in improving resource scaling efficiency.
Fig.~\ref{fig:example_cpu} shows the CPU core hour usage of each microservice under different strategies. We can observe that microservices $C$, $F$, $H$ have significant resource usage differences under different strategies. Specially, for the microservice $C$, MAX-Scale uses 2.2X CPU core hours than FineGrained-Scale. 
Looking into call graphs related to microservice $C$ shown in Fig.~\ref{fig:cg_example}, we can observe $H$ calls $C$ on {\it func1} with one time in {\it cg1}, calls on {\it func1} with two times in {\it cg2}, and calls {\it func1}, {\it func2}, and {\it func3} once each in {\it cg3}.
The repeated calls and different interfaces result in significant differences in resource usage patterns among {\it cg1}, {\it cg2}, and {\it cg3}. 
For example, from our profiling at the same load, the resource demand of {\it cg3} is 5.2X of {\it cg1}.

FineGrained-Scale can accurately identify call graphs with fine-grained call graph constructing, while MAX-Scale and MIN-Scale coarsely map these call graphs to {\it coarse-grained cg} (Fig.~\ref{fig:cg_example}(d)).
In this case, FineGrained-Scale can accurately determine the just-enough resources for the microservices. 
By contrast, MAX-Scale allocates resources for microservices according to the maximum resource usage ({\it func3}) and repeated calling times (3 times) per query. MIN-Scale allocates the minimum resource ({\it func1} with one time ) for microservices, which results in the QoS violations as shown in Fig.~\ref{fig:example_latency}.




\subsubsection{Overhead of the FineGrained-Scale}
For each query, FineGrained-Scale's call graph constructing involves traversing each microservice call and updating the weight of the corresponding edge. The time complexity is O(m), where m is the number of inter-microservice calls triggered by a query. For constructing N queries during a monitoring interval, the time complexity is O(Nm).
In the Alibaba trace, the query number per minute for a single service does not exceed 30,000. 
We evaluate FineGrained-Scale with this maximum possible load, and the total time is approximately 2100ms. This is acceptable relative to the one-minute monitoring interval.

\section{Conclusion\label{sec:conclusion}}

This paper proposes DGG for generating service dependency graphs of benchmarks that incorporate production-level features. 
Specifically, DGG uses a data handler to construct precise call graphs from the production traces and merges them into dependency graphs. It then clusters these dependency graphs into different categories based on their topological and invocation types. 
Based on the organized data, DGG uses a graph generator to generate service dependency graphs based on the random graph models that simulate real microservices invoking downstream microservices.
Results show that DGG's generated dependency graphs closely resemble real traces. Moreover, resource management based on DGG's fine-grained call graph constructing increases resource efficiency of microservices by up to 44.8\% while ensuring the QoS.

\bibliographystyle{IEEEtran}
\bibliography{IEEEabrv,reference}

\end{document}